\documentclass{aastex}
\usepackage{emulateapj5}
\usepackage{onecolfloat}
\usepackage{apjfonts}

\newcommand{\ob}{\omega_{\rm b}}
\newcommand{\oc}{\omega_{\rm cdm}}
\newcommand{\Ob}{\Omega_{\rm b}}
\newcommand{\Oc}{\Omega_{\rm cdm}}
\newcommand{\Oq}{\Omega_{\rm Q}}
\newcommand{\wq}{w_{\rm Q}}

\begin{document}
\twocolumn[%

\submitted{Accepted by ApJL}

\title{Probing dark energy with the cosmic microwave background:
  projected constraints from the {\it Wilkinson Microwave Anisotropy
    Probe\/} and {\it Planck\/}}

\author{A. Balbi\altaffilmark{1}, C. Baccigalupi\altaffilmark{2}, F.
  Perrotta\altaffilmark{2}, S. Matarrese\altaffilmark{3}, N.
  Vittorio\altaffilmark{1}}

\begin{abstract}
  We investigate the accuracy attainable by forthcoming space-based
  observations of the cosmic microwave background (CMB) temperature
  and polarization anisotropy in constraining the dark energy density
  parameter $\Oq$ and equation of state $\wq=p_{\rm Q}/\rho_{\rm Q}$.
  Despite degeneracies among parameters, it is possible for high
  precision observations such as those from the {\it Wilkinson
    Microwave Anisotropy Probe\/} and {\it Planck\/} to provide
  interesting information on the nature of the dark energy.
  Furthermore, we show that imposing a flat universe constraint makes
  it possible to obtain tight limits in the space of dark energy
  parameters even from the CMB alone.
\end{abstract}


] 
\altaffiltext{1}{Dipartimento di Fisica, Universit\`a Tor Vergata,
  and INFN, Sezione di Roma 2, Via della Ricerca Scientifica 1, 00133
  Roma Italy; balbi@roma2.infn.it, vittorio@roma2.infn.it.}
\altaffiltext{2}{SISSA/ISAS, Via Beirut 4, 34014 Trieste, Italy;
  bacci@sissa.it, perrotta@sissa.it.}  
\altaffiltext{3}{Dipartimento di Fisica `Galileo Galilei',
  Universit\'a di Padova, and INFN, Sezione di Padova, Via Marzolo 8,
  35131 Padova, Italy; matarrese@pd.infn.it.}

\section{Introduction}
Recent cosmological observations have shed new light on the cosmic
budget problem. There is now strong evidence from CMB anisotropy
measurements (see e.g., Benoit et al.~2002 and references therein)
that the universe has a total energy density close to critical (and
therefore a flat large scale geometry) and that matter (either
luminous or dark) can only account for about $30\%$ of the total
density (see e.g., Turner 2002). On the other hand, direct evidence
for cosmic acceleration from high redshift type Ia supernovae
observations (Riess et al.~1998; Perlmutter et al.~1999) can be
interpreted as indication for the existence of a smooth dark energy
component with equation of state $\wq\equiv p_{\rm Q}/\rho_{\rm Q} <
-1/3$, that would be responsible for the remaining $70\%$ of the
critical density. The nature of this dark energy component, however,
remains a mystery. In particular, little is known about its equation
of state. The simplest possible kind of dark energy is the vacuum
energy (or cosmological constant), with $\wq=-1$ independent of time.
However, any plausible scalar field from fundamental theories has a
vacuum expectation value that would overclose the universe by tens of
orders of magnitude. More general scalar fields, termed quintessence,
may be spatially inhomogeneous, with $\wq\neq -1$ and varying in time.
A number of strategies have been proposed to investigate the nature of
dark energy: supernovae observations at multiple redshifts (Huterer \&
Turner 2001), weak lensing (Huterer 2002), cluster counting (Haiman,
Mohr \& Holder 2001), redshift surveys (Matsubara \& Szalay 2003),
Lyman-$\alpha$ forest (Viel et al.\ 2003), CMB anisotropy (Baccigalupi
et al.\ 2002), as well as combinations of these methods (see, e.g.,
Kujat et al.\ 2002 and references therein).

In this work, we investigate the ability of space-based high-resolution
CMB anisotropy observations to constrain the dark energy parameters.
We focus on the currently underway {\em Wilkinson Microwave Anisotropy
  Probe (WMAP)\/}\footnote{See http://map.gsfc.nasa.gov.}  satellite
mission and on the forthcoming {\em Planck Surveyor\/}\footnote{See
  http://astro.estec.esa.nl/Planck.} and we produce projections for the
model parameter errors that are attainable by these experiments.  We model the
dark energy component as a minimally coupled quintessence field
following tracking trajectories for an inverse power law potential
(Ratra \& Peebles 1988; Wetterich 1988).  Such a field is in general
characterized by two main features: $(1)$ spatial inhomogeneities are
on scales comparable to the horizon (whereas in non-minimally coupled
theories they can generally appear on all scales [Perrotta \&
Baccigalupi 2002]) and $(2)$ the equation of state does not vary in time
during the tracking regime, when quintessence is subdominant; in this
regime the equation of state is fixed by the attractor solution of the
Klein Gordon equation. The equation of state can, however, change
significantly at low redshifts, when quintessence is no longer
subdominant, and in this case it tends to a cosmological constant
behavior, thus having a present equation of state generally {\it
  smaller} than during the tracking regime.

\section{Method}
To quantify how well {\em WMAP\/} and {\em Planck\/} can estimate the
dark energy parameters, we followed a Fisher information matrix
approach (Fisher 1935; see also Tegmark, Taylor \& Heavens 1997). The
Fisher information matrix is defined as the expectation value
\begin{equation}
F_{ij}\equiv - \left\langle{\partial^2\log L \over \partial p_i\ \partial p_j}
\right\rangle
\end{equation}
where $L$ is the likelihood of the data and $p_i$ are the model
parameters. The diagonal elements of the inverse Fisher matrix give
the minimum variance of model parameters estimated from a given
dataset, assuming an underlying fiducial ``target'' model. This
approach relies on the assumption that the likelihood is well
approximated by a Gaussian around its peak.

\begin{table*}[t!]
  \caption{Experimental Parameters\label{tab:experiments}}
  \tablecomments{Sensitivities to temperature and polarization, 
    $\sigma^T$ and $\sigma^P$, are relative to the average CMB 
    temperature (2.73 K). A pixel is a square whose side is the FWHM 
    extent of the beam.}
  \begin{center}
    \begin{tabular}{lrrrrrrrrrrrrr}
      \tableline
      \tableline
       & \multicolumn{3}{c}{{\em WMAP\/}} & & \multicolumn{3}{c}{{\em Planck\/}/LFI} & & \multicolumn{4}{c}{{\em Planck\/}/HFI}\\
      \tableline
      Center frequency (GHz) \dotfill  & 
      40 & 60 & 90 & & 44 & 70 & 100 & & 100 & 143 & 217 & 353 \\
      Angular resolution (FWHM, arcmin.) \dotfill  & 
      31.8 & 21 & 13.8 & & 24 & 14 & 10 & & 9.2 & 7.1 & 5.0 & 5.0 \\
      $\sigma^T$ per pixel\ $(\times 10^{-6})$ \dotfill & 
      4.1 & 9.4 & 21.8 & & 2.7 & 4.7 & 6.6 & & 2.0 & 2.2 & 4.8 & 14.7 \\
      $\sigma^P$ per pixel\ $(\times 10^{-6})$ \dotfill & 
      \nodata & \nodata & \nodata & & 3.9 & 6.7 & 9.3 & & \nodata & 4.2 & 9.8 & 29.8 \\
      \tableline
    \end{tabular}
  \end{center}
\vspace{-1cm}
\end{table*}

The Fisher matrix depends on the cosmological model and parameters
adopted, as well as on the covariance matrix of the data, which in
turn depends on the experimental setup. 
When only temperature information is considered, the Fisher matrix can
be calculated from the theoretical power spectrum of CMB anisotropy
$C_l$ and its first derivatives with respect to cosmological
parameters $p_i$, according to the formula:
\begin{equation}\label{fisher}
F_{ij} = \sum_l {\partial C\over \partial p_i}
         {1\over (\Delta C_l)^2}{\partial C\over \partial p_j}.
\end{equation}
For an experiment sampling a fraction of sky $f_{\rm sky}$ with $N$
frequency channels $c$, each with average sensitivity per pixel
$\sigma_c$ and a Gaussian beam response $B_{c,l}$, with FWHM angular
resolution $\theta_c$, if we define $w\equiv\sum_c w_c\equiv\sum_c
(\sigma_c\theta_c)^{-2}$ and $B_l^2\equiv \sum B_{c,l}^2 w_c/w$, we can
write (see Knox 1995):
\begin{equation}
(\Delta C_l)^2\approx {2\over (2l+1)f_{\rm sky}}
                      \left(C_l+w^{-1} B_l^{-2}\right)^2
\end{equation}
When the polarization information is also considered (as in this
Letter), then equation (\ref{fisher}) has to include the full covariance
matrix of the $C_l$ for all the temperature and polarization
components: the full formalism, which we adopt in our analysis, is
given, for example, by Zaldarriaga, Spergel \& Seljak (1997).

The {\em WMAP\/} and {\em Planck\/} experimental parameters assumed in
our analysis are summarized in Table~\ref{tab:experiments}.  We
included in the analysis three frequency channels from {\em WMAP\/},
three from {\em Planck\/} Low Frequency Instrument (LFI) and four from
{\em Planck\/} High Frequency Instrument (HFI). We fully took into
account the {\em Planck\/} polarization capabilities. For both
experiments, we conservatively assumed that foreground contamination
from the Galactic plane leaves unobserved a symmetric strip of $\pm
20^\circ$ around the Galactic equator, so that the observed fraction
of the sky is $f_{\rm sky}=0.66$.
The mission duration required to meet the experimental specifications
used in this paper is 2 years of continuous observation for {\em
  WMAP\/} and 14 months (two sky surveys) for {\em Planck\/}.

We considered inflationary adiabatic cosmological models with eight
free parameters: the present-day dark energy density $\Oq$ and
equation of state $\wq$; the total energy density of the universe
$\Omega$; the physical baryon and cold dark matter densities
$\ob\equiv \Ob h^2$ and $\oc \equiv \Oc h^2$ (here $h$ is the present
value of the Hubble parameter in units of 100~km/s/Mpc); the
primordial spectral index of scalar perturbations $n_s$; the overall
amplitude of the CMB power spectrum in units of the {\em
  COBE\/}-normalized $C_{10}$ multipole; the ratio $R$ between the
tensor and scalar contribution to the CMB quadrupole.  We used a
single-field inflation consistency relations between the tensor
amplitude and spectral index: $n_T=-R/6.8$.  We assumed no massive
neutrino contribution and a negligible reionization optical depth.
Note that $\Omega=\Omega_{\rm M}+\Oq$, where $\Omega_{\rm M}=\Ob+\Oc$
is the total matter density: therefore $h$ is a dependent parameter,
determined by the constraint: $h=[(\ob+\oc)/(\Omega-\Oq)]^{1/2}$. The
choice of $\Omega$, $\ob$ and $\oc$ as free parameters (rather than,
for example, $\Ob$, $\Oc$ and $h$) has become usual practice in this
sort of analysis. In fact, the combinations $\ob$ and $\oc$ directly
govern the physics of acoustic oscillations which defines the CMB
anisotropy pattern, while $\Omega$ fixes the geometry of the universe
and then the angular size of characteristic features on the CMB. For
this reason, $\Omega$, $\ob$ and $\oc$ are much better constrained by
the CMB than other combinations of parameters, and are therefore a
more suitable choice in a Fisher matrix analysis (Efstathiou \& Bond
1999).

We chose our target to be the flat quintessence model which best fits
the currently available CMB data (Baccigalupi et al.\ 2002).  The
parameters of this model are summarized in Table~\ref{tab:target}.  To
quantify variations around the target model we numerically computed
two-sided derivatives of the theoretical CMB angular power spectrum
with respect to the parameters, using a step size which was roughly
5\% of the target parameter value.
All the theoretical spectra were computed using a modified version of
CMBFAST (Seljak \& Zaldarriaga 1996).
The variation of the total energy density was taken into account
through its effect on the angular diameter distance, which, for fixed
$\ob$ and $\oc$, just results in a shift in multipole space of the CMB
angular power spectrum.

\begin{table}[tb]
  \caption{Target model\label{tab:target}} 
  \tablecomments{$C_{10}$ is normalized to COBE.}
  \begin{center}
    \begin{tabular}{ll}
      \tableline
      \tableline
      Parameter & Value \\
      \tableline
      $\wq$ \dotfill                  & $-0.8$      \\
      $\Oq$ \dotfill                  & \phs$0.7$   \\
      $\Omega$ \dotfill               & \phs$1$     \\
      $\ob$ \dotfill                  & \phs$0.022$ \\
      $\oc$ \dotfill                  & \phs$0.145$ \\
      $n_s$ \dotfill                  & \phs$1$     \\
      $R$ \dotfill                    & \phs$0.1$   \\
      $h$ \dotfill                    & \phs$0.746$ \\
      \tableline
    \end{tabular}
  \end{center}
\end{table}

\section{Results}

\begin{figure}
  \plotone{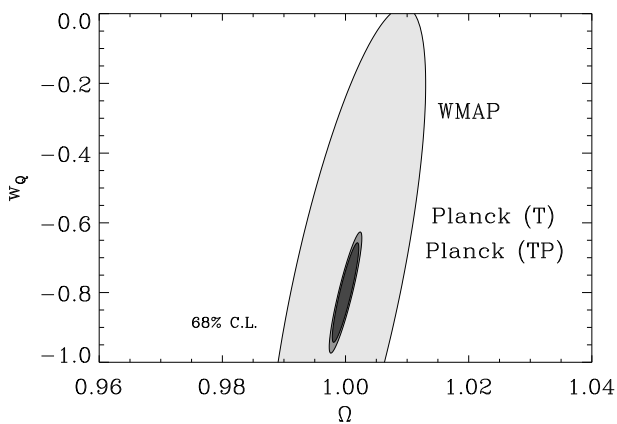}
  \caption{The 68\% confidence level constraints in the ($\Omega$, $\wq$) 
    parameter space. The shaded regions ({\em lighter to darker}) are
    obtained from {\em WMAP\/} and {\em Planck\/} (temperature only)
    and from {\em Planck\/} (temperature and polarization).
    \label{fig:wqomega}}
\end{figure}

The main effect of a dark energy component on the CMB anisotropy
pattern is purely geometric. Varying the dark energy equation of state
changes the angular diameter distance (by changing the expansion rate
of the universe), resulting in a shift of features in the angular
power spectrum of the CMB towards larger angular scales (lower
multipoles) as $\wq$ gets larger than $-1$. Varying the total energy
density of the universe, $\Omega$, also changes the angular diameter
distance, because of the geodesic deviation of CMB photons from
recombination to the present. We thus expect to observe a degeneracy
between $\wq$ and $\Omega$, i.e. a variation in the angular diameter
distance due to $\wq$ can be compensated by an opposite variation due
to $\Omega$.  This is just an aspect of the well-known geometrical
degeneracy (Bond, Efstathiou \& Tegmark 1997; Zaldarriaga, Spergel \&
Seljak, 1997) inherent in any CMB anisotropy measurement. The
degeneracy is not exact because different values of $\wq$ result in a
different integrated Sachs-Wolfe contribution at large angular scales:
however, this is precisely where the cosmic variance uncertainty on
the CMB angular power spectrum is larger.  The amount of degeneracy
can be quantified by investigating the covariance between $\wq$ and
$\Omega$, obtained from the 2x2 submatrix of $F^{-1}$ corresponding to
this pair of parameters. In Figure~\ref{fig:wqomega} we show the 68\%
confidence level constraints in the ($\Omega$, $\wq$) plane obtained
with this method. Clearly, varying $\wq$ even by a considerable amount
has a much weaker effect than varying $\Omega$. As a result, the dark
energy equation of state is poorly determined by CMB observations,
even though the high sensitivity achievable by {\em Planck\/} allows one to
put an upper limit to $\wq$. On the other hand, the determination of
$\Omega$ is not very much affected by variations in the dark energy
equation of state, because of the much stronger dependence of the
angular diameter distance on $\Omega$.  Note that these results do not
change when CMB polarization is included in the analysis.

\begin{figure*}[tb]
  \epsscale{2.2}
  \plottwo{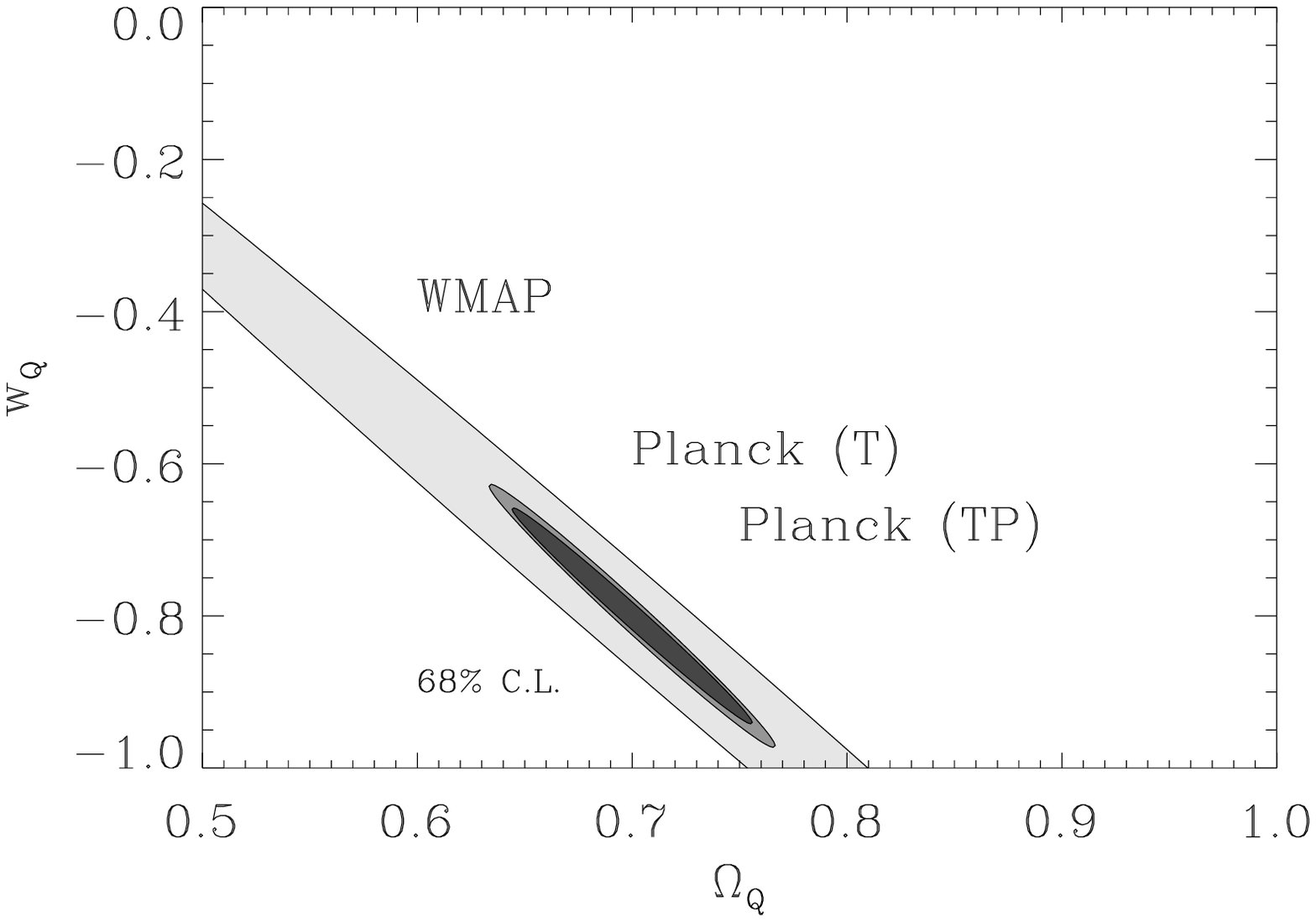}{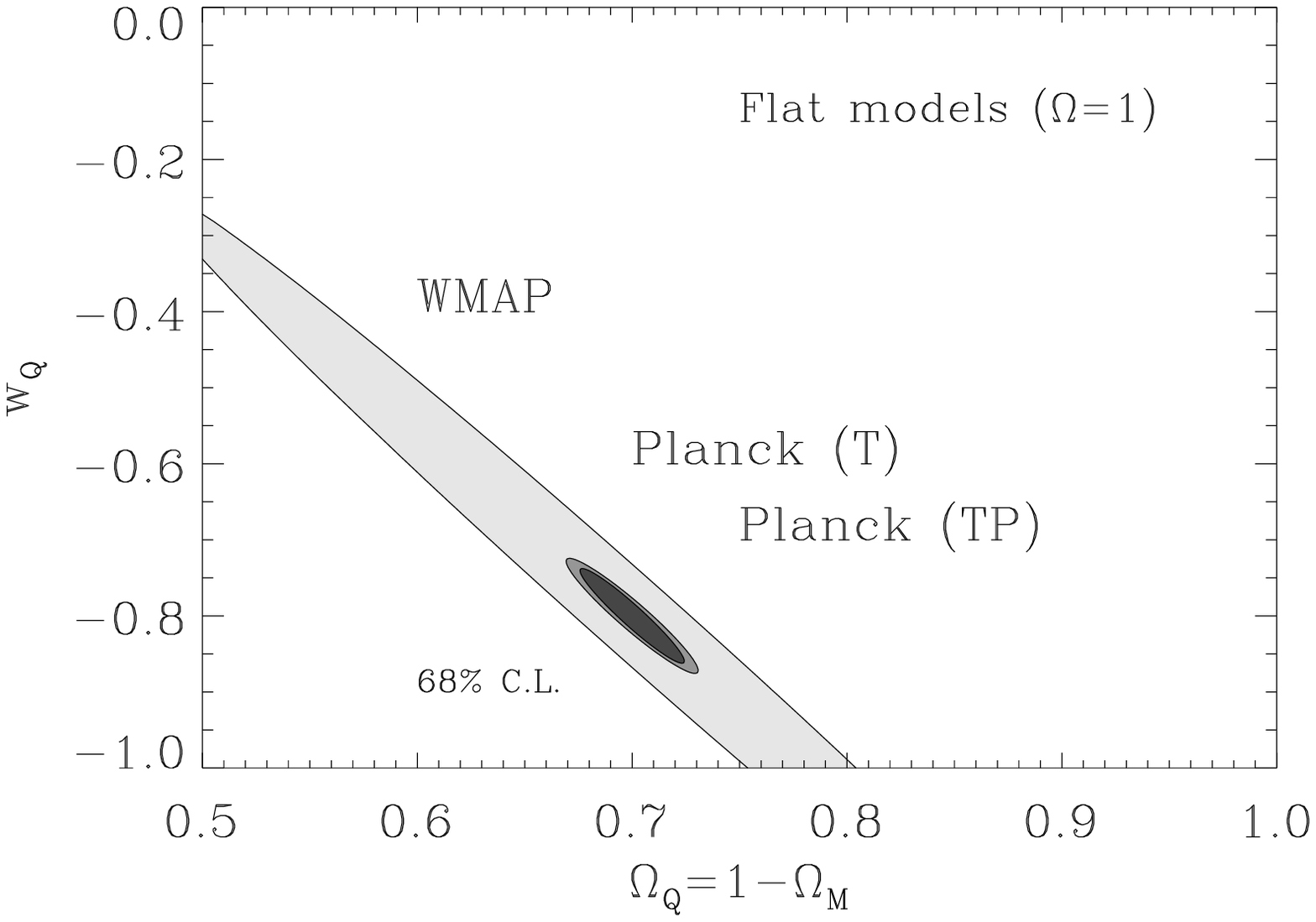}
  \caption{The 68\% confidence level constraints in the dark energy 
    parameters space ($\Oq$, $\wq$). {\em Left}: Shaded regions ({\em
      lighter to darker}) obtained from {\em WMAP\/} and {\em
      Planck\/} (temperature only) and from {\em Planck\/}
    (temperature and polarization). {\em Right}: same contours, except
    we impose the flat universe constraint
    ($\Omega=1$).\label{fig:wqomegaq}}
\end{figure*}

Constraints in the space of dark energy parameters ($\Oq$, $\wq$)
obtained with the same technique are shown in Figure
\ref{fig:wqomegaq}. Again, there exists a strong degeneracy between
the two parameters: due to this, {\em WMAP\/} is basically unable to
distinguish our target model from a cosmological constant case
$\wq=-1$.  The situation improves when a flat universe ($\Omega=1$) is
assumed. This additional constraint partially breaks the degeneracy,
reducing the allowed region in the dark energy parameter space. The
improvement is dramatic for {\em Planck\/}: the confidence level contours
get closed, enabling an accurate determination of the dark energy
parameters.

Our results are summarized in Table \ref{tab:errors}, where we show
the projected 1$\sigma$ error bars on each parameter of our model when
marginalizing the others. As it is well known, some cosmological
parameters are constrained to very high accuracy by the CMB: notably,
the baryon and cold dark matter physical densities, which control the
relative peak heights in the angular power spectrum, can be measured
to an accuracy of better than percent. Geometrical effects do not
change this result considerably, which therefore remains valid when
different dark energy equations of state are considered. Even in the
presence of a strong geometrical degeneracy, {\em Planck\/} is able to
set interesting constraints on the dark energy equation of state: our
target model $\wq=-0.8$ can be distinguished at 1$\sigma$ from the
cosmological constant case $\wq=-1$ using {\em Planck\/} measurements.
Imposing the flat universe constraint $\Omega=1$ results in a decrease
of roughly a factor of 2 in the error bars for the dark energy
parameters $\wq$ and $\Oq$.

\begin{table*}[tb]
  \caption{Marginalized 1$\sigma$ errors\label{tab:errors}}
  \tablecomments{Errors relative to the target model of Table \ref{tab:target}. 
    The number in parentheses were obtained by imposing the additional 
    constraint $\Omega=1$.} 
  \begin{center}
    \begin{tabular}{llll}
      \tableline
      \tableline
      Parameter & {\em WMAP\/} & {\em Planck\/}, T only & {\em Planck\/}, T and P \\
      \tableline
      $\Delta \wq$ \dotfill           
      & 0.54 (0.37) & 0.11 (0.050) & 0.095 (0.041) \\
      $\Delta \Oq$  \dotfill     
      & 0.22 (0.15) & 0.044 (0.020) & 0.037 (0.016) \\
      \tableline
      $\Delta \Omega$  \dotfill       
      & 0.0086 & 0.0017  & 0.0014 \\
      $\Delta \ob$ \dotfill           
      & 0.00039 (0.00038) & $9.6\times 10^{-5}$ ($9.6\times 10^{-5}$) & $8.2\times 10^{-5}$ ($8.2\times 10^{-5}$) \\
      $\Delta \oc$ \dotfill           
      & 0.0041 (0.0040) & 0.0018 (0.0015) & 0.0012 (0.0011) \\
      $\Delta n_s$ \dotfill           
      & 0.011 (0.011) & 0.0042 (0.0034) & 0.0031 (0.0027) \\
      $\Delta C_{10}/C_{10}$ \dotfill 
      & 0.044 (0.044) & 0.041 (0.041) & 0.020 (0.020) \\
      $\Delta R$ \dotfill             
      & 0.071 (0.071) & 0.054 (0.053) & 0.025 (0.024) \\
      $\Delta h$ \dotfill 
      & 0.29 (0.20)   & 0.061 (0.028) & 0.051 (0.022) \\
      \tableline
    \end{tabular}
  \end{center}
\end{table*}

\section{Discussion}
In this Letter, we quantified the capability of {\em WMAP\/} and {\em
  Planck\/} to constrain dark energy cosmologies. The issue of the
importance of CMB measurements for dark energy has recently received
considerable attention and different interpretations: for example,
Baccigalupi et al.\ (2002) have shown that with a strong prior on the
Hubble constant one could constrain the dark energy at the level of
$10\%$ even using present CMB data, also obtaining an indication that
$w_{Q}\simeq -0.8$. On the other hand, Bean \& Melchiorri (2002) have
shown that when that prior is relaxed, our knowledge of dark energy
from CMB data alone is still poor.

For the background cosmology, we evaluated the attainable precision of
measurements of the dark energy abundance and equation of state, the
amount of baryons and cold dark matter, the Hubble constant and the
cosmological curvature; for the early universe, we considered the
perturbation normalization, the scalar spectral index, and the ratio
between tensor and scalar perturbations, by only assuming single-field
inflation consistency relations between the tensor amplitude and
spectral index. We found that the performance of {\em Planck\/} is a factor 4
to 6 better than {\em WMAP\/} in all cases. For the tensor-to-scalar ratio, the
inclusion of polarization for {\em Planck\/} allows to gain a factor 2 in
accuracy, as expected since polarization is directly susceptible to
tensor perturbations.

As for the dark energy, we confirmed expected parameter degeneracies:
indeed, the main effect of $\wq>-1$ on the CMB is to decrease the
distance to last scattering; however, such effect can be mimicked by a
closed geometry (i.e.\ $\Omega>1$).  Inclusion of polarization
information does not help breaking these degeneracies. On the other
hand, despite such drawbacks, our main result is that the capability
of {\em Planck\/} to constrain the dark energy is quite good; this
statement holds in particular when we compare the {\em
  Planck\/}forecasted constraints on the dark energy parameters with
those from dedicated experiments such as the {\em
  Supernova/Acceleration Probe (SNAP)\/}\footnote{See
  http://snap.lbl.gov.}.  The latter is expected to reach a precision
of a few percent in both dark energy abundance and equation of state.
According to our results, {\em Planck\/} should perform at the level
of a few percent accuracy on the dark energy abundance, but of only
about $10\%$ on the equation of state, mainly because of the
degeneracy with the spatial curvature; if the latter is set to zero
(i.e.\ if $\Omega=1$), the equation of state can be determined at the
level of about $5\%$.  We also recall that, as it is well known,
contours in the ($\Oq$, $\wq$) plane obtained from supernova
measurements are roughly orthogonal to those obtained from the CMB,
thus maximizing the accuracy that can be achieved by combining the two
methods of observation. Due to the precision attainable by {\em
  Planck\/} and {\em SNAP}, it is reasonable to expect that an
accuracy of about $1\%$ on the determination of the dark energy
parameters will be obtained by complementing the results of the two
experiments.

We point out that the results obtained here do not depend
significantly on the target model assumed or on the quintessence
details; indeed, we repeated the same analysis by adopting an
effective dark energy model, having a constant equation of state, and
another target model, with equation of state $\wq=-1$, obtaining
similar results.

We conclude that the CMB is one of the most sensitive observables to
the main dark energy parameters, i.e.\ its abundance and equation of
state.
In fact, the recently released results from {\em WMAP\/}'s first year
of observation (Bennett et al.~2003) are encouraging in this sense
(Spergel et al.~2003): the actual parameter constraints are somewhat
broader than estimated in our forecast, but they should get tighter
when more data are collected.
By combining CMB observations with those from dedicated experiments
such as {\em SNAP\/}, we can expect to measure the equation of state
at the level of percent.  Finally we also recall that the use of CMB
for any cosmological measurement, including dark energy, has the great
advantage that CMB perturbations are linear and therefore relatively
simple to describe.  With any other observable from large-scale
structure, we need to fully understand the non-linear structure
formation in dark energy cosmology.

\acknowledgements{The authors would like to thank Lloyd Knox and Paolo
  Natoli for useful discussions.}


\begin{thebibliography}{99}
\bibitem[Baccigalupi et al.]{BBPMV02} Baccigalupi, C., Balbi, A.,
  Matarrese, S., Perrotta, Vittorio, N. 2002 \prd 65, 063520
\bibitem[Bennett et al.]{Bennett} Bennett, C.L., et al. 2003, 
submitted to \apj\ [astro-ph/0302207]
\bibitem[Benoit et al.]{archeops} Benoit, A., et al. 2003, A\&A, 399, L25
\bibitem[Bean \& Melchiorri]{Bean} Bean, R., \& Melchiorri, A. 2002,
  \prd 65, 041302
\bibitem[Bond, Efstathiou \& Tegmark]{BET97} Bond, J.R., Efstathiou,
  G. \& Tegmark, M. 1997, \mnras, 291, 33
\bibitem[Efstathiou \& Bond(1999)]{EB} Efstathiou, G.~\& Bond, J.~R.\ 
  1999, \mnras, 304, 75
\bibitem[Fisher]{Fisher35} Fisher, R.A. 1935, J.Roy.Stat.Soc., 98, 39
\bibitem[Haiman, Mohr, \& Holder(2001)]{2001ApJ...553..545H} Haiman,
  Z., Mohr, J.~J., \& Holder, G.~P.\ 2001, \apj, 553, 545
\bibitem[Huterer]{Huterer} Huterer, D. 2002, \prd 65, 063001
\bibitem[Huterer \& Turner(2001)]{2001PhRvD..64l3527H} Huterer, D.~\&
  Turner, M.~S.\ 2001, \prd 64, 123527
\bibitem[Knox]{Knox95} Knox, L. 1995, \prd 52, 4307 
\bibitem[Kujat et al.]{Kujat} Kujat, J., Linn, A.M., Scherrer, R.J., 
Weinberg, D.H., 2002, \apj, 572, 1
\bibitem[Matsubara]{MS} Matsubara, T. \& Szalay, A.S. 2003, \prl\ 90, 021302
\bibitem[Perlmutter]{SN2} Perlmutter, S. et al. 1999, \apj, 517, 565
\bibitem[Perrotta \& Baccigalupi]{PB} Perrotta, F. \& Baccigalupi, C.
  2002, \prd 65, 123505
\bibitem[Ratra \& Peebles]{RP} Ratra, B. \& Peebles, P.J.E. 1988, \prd
  37, 3406
\bibitem[Riess]{SN1} Riess, A. et al. 1998, \aj, 116, 1009
\bibitem[cmbfast]{cmbfast} Seljak, U., \& Zaldarriaga, M. 1996, \apj,
  469, 437
\bibitem[Spergel]{Spergel} Spergel, D.N., et al. 2004, submitted to
  \apj\ [astro-ph/0302209]
\bibitem[Tegmark, Taylor \& Heavens]{TTH97} Tegmark, M., Taylor, A.N.,
  \& Heavens, A.F. 1997, \apj, 480, 22
\bibitem[Turner]{Turner02} Turner, M.S. 2002 \apj, in press
  [astro-ph/0106035]
\bibitem[Viel]{Viel} Viel, M., Matarrese, S., Theuns, T., Munshi D.,
  Wang, Y., 2003, \mnras, in press [astro-ph/0212241]
\bibitem[Wetterich]{W} Wetterich, C. 1988, Nucl. Phys. B 302 668
\bibitem[Zaldarriaga, Spergel \& Seljak]{ZSS97} Zaldarriaga, M.,
  Spergel, D.N. \& Seljak, U. 1997, \apj, 488, 1
\end{thebibliography}
\end{document}